\documentclass[11pt,a4paper]{article}
\usepackage{amsmath,amssymb,mathrsfs}
\usepackage[dvipdfmx]{graphicx}

\def\un#1{\relax\ifmmode\@@underline#1\else
        $\@@underline{\hbox{#1}}$\relax\fi}

\let\du=\du                     

\def\a{\alpha}
\def\b{\beta}

\def\f{\phi}
\def\g{\gamma}

\def\m{\mu}
\def\n{\nu}

\def\z{\zeta}

\def\F{\Phi}

\def\L{\Lambda}

\def\ve{\varepsilon}

\def\ce{{\cal E}}

\def\car{{\cal R}}

\def\ct{{\cal T}}

\def\cy{{\cal Y}}

\def\bo{{\raise-.3ex\hbox{\large$\Box$}}}               
\def\pa{\partial}                                       
\def\TH{{\raise.2ex\hbox{$\displaystyle \bigodot$}\mskip-4.7mu \llap H \;}}
\def\face{{\raise.2ex\hbox{$\displaystyle \bigodot$}\mskip-2.2mu \llap {$\ddot
        \smile$}}}                                      

\def\Bar#1{\overline{#1}}                       
\def\leftrightarrowfill{$\mathsurround=0pt \mathord\leftarrow \mkern-6mu
        \cleaders\hbox{$\mkern-2mu \mathord- \mkern-2mu$}\hfill
        \mkern-6mu \mathord\rightarrow$}
\def\dvec#1{\vbox{\ialign{##\crcr
        \leftrightarrowfill\crcr\noalign{\kern-1pt\nointerlineskip}
        $\hfil\displaystyle{#1}\hfil$\crcr}}}           
\def\dt#1{{\buildrel {\hbox{\LARGE .}} \over {#1}}}     

\def\sfrac#1#2{{\vphantom1\smash{\lower.5ex\hbox{\small$#1$}}\over
        \vphantom1\smash{\raise.4ex\hbox{\small$#2$}}}} 
\def\bfrac#1#2{{\vphantom1\smash{\lower.5ex\hbox{$#1$}}\over
        \vphantom1\smash{\raise.3ex\hbox{$#2$}}}}       
\def\afrac#1#2{{\vphantom1\smash{\lower.5ex\hbox{$#1$}}\over#2}}    

\def\[{\lfloor{\hskip 0.35pt}\!\!\!\lceil}
\def\]{\rfloor{\hskip 0.35pt}\!\!\!\rceil}

\def\du#1#2{_{#1}{}^{#2}}

\def\ha{{\fracmm12}}

\def\un{\underline}
\def\fracmm#1#2{{{#1}\over{#2}}}

\def\low#1{{\raise -3pt\hbox{${\hskip 0.75pt}\!_{#1}$}}}

\def\Dot#1{\buildrel{_{_{\hskip 0.01in}\bullet}}\over{#1}}
\def\dt#1{\Dot{#1}}

\newskip\humongous \humongous=0pt plus 1000pt minus 1000pt
\def\caja{\mathsurround=0pt}
\def\eqalign#1{\,\vcenter{\openup2\jot \caja
        \ialign{\strut \hfil$\displaystyle{##}$&$
        \displaystyle{{}##}$\hfil\crcr#1\crcr}}\,}
\newif\ifdtup

\newcommand{\be}{\begin{equation}}
\newcommand{\ee}{\end{equation}}
\newcommand{\nbe}{\begin{equation*}}
\newcommand{\nee}{\end{equation*}}

\newcommand{\lb}{\label}

\begin{document}

\begin{titlepage}

\begin{center}

October 2013 \hfill IPMU13-0169\\
revised version 

\noindent
\vskip2.0cm
{\Large \bf 

On the Supersymmetrization of Inflation in $f(R)$ Gravity 

}

\vglue.3in

{\large
Sergei V. Ketov~${}^{a,b,c}$ 
}

\vglue.1in

{\em
${}^a$~Department of Physics, Tokyo Metropolitan University \\
Minami-ohsawa 1-1, Hachioji-shi, Tokyo 192-0397, Japan \\
${}^b$~Kavli Institute for the Physics and Mathematics of the Universe (IPMU)
\\The University of Tokyo, Chiba 277-8568, Japan \\
${}^c$~Department of Physics, University of Oslo, \\ 
P.B. 1048, Blindern, 0315 Oslo, Norway
}

\vglue.1in
ketov@phys.se.tmu.ac.jp

\vglue.3in

\end{center}

\begin{abstract}
An N=1 Poincar\'e supergravity action, suitable for describing the Starobinsky inflation, is proposed.
It extends $f(R)$ gravity to supergravity in its old-minimal version. The action is parametrized 
by a single holomorphic potential and a single non-holomorphic potential, and can be dualized 
into the standard matter-coupled supergravity action, with the "matter" given by two chiral superfields. 
The action extends the earlier proposals for embedding the Starobinsky inflation to supergravity, and can be 
further generalized to describe quantum corrections to the inflation. Our approach assumes the gravitational 
origin of inflaton and quintessence in the context  of supergravity, by using a single chiral scalar curvature superfield.
\end{abstract}

\end{titlepage}


\section{Introduction}\label{sec:Intro}

The most recent (March 2013) PLANCK satellite mission data \cite{planck} combined with the WMAP9 polarization and lensing data \cite{wmap} yield $n_s = 0.960 \pm 0.007$ for the CMB spectral index, $r<0.08$ for the CMB tensor-to-scalar-ratio with the 95\% level of confidence, and $f_{NL} = 2.7 \pm 5.8$  with the 68\% level of confidence for the CMB non-Gaussianity parameter. This data apparently favors the inflationary models with relatively low $r$ and low non-Gaussianity. The Starobinsky inflationary model based on the $(R+R^2)$ gravity \cite{star} perfectly fits the bill, so it inspired renewed interest to that model in the recent literature in the context of supergravity (see eg. Ref.~\cite{fklp} and references therein). Embedding the Starobinsky inflation into supergravity is the important step towards embedding inflation into unified theories of particle physics, including superstrings.

 The simplest Starobinsky model is described by the action (see Ref.~\cite{myrev} for our
notation with the reduced Planck mass $M_{\rm Pl}=1$) 
\be \lb {stara}
S[g] =-\ha \int d^4x\sqrt{-g} \left[ R -R^2/(6M^2)\right]
\ee
in terms of metric $g_{\m\n}(x)$ having the scalar curvature $R$, with the only mass
parameter $M$. The parameter $M$ is fixed by the CMB data as $M=(3.0 \times
10^{-6})(\fracmm{50}{N_e})$ where $N_e$ is the e-foldings number. The action (\ref{stara}) can be dualized by the Legendre-Weyl transform \cite{lwtr}  to the standard quintessence acton with the celebrated scalar potential
\be \lb{starp}
V(\f) = \fracmm{3}{4} M^2\left( 1- e^{-\sqrt{\frac{2}{3}}\f }\right)^2
\ee
that is quite suitable for viable inflation at large positive $\f$ rolling down over the plateau.
The physical meaning of inflaton (dubbed {\it scalaron}) with the mass M in the Starobinsky model is given by the spin-0 part of metric and thus has the clear geometrical origin. The gravitational origin of scalaron is obscure in the quintessence picture. Knowing the inflaton origin is essential for fixing its interactions with other (matter) fields, which is instrumental for reheating after inflation \cite{myrev}.

Some comments are in order here. First, the action (\ref{stara}) has higher derivatives but the equivalent quintessence action has not (of course, this equivalence is classical). It means that the ghosts can be avoided in the theory (\ref{stara}) despite of the presence of the higher derivatives. Second, it demonstrates the "accidental" drop of the spin-0 part of metric in the Einstein-Hilbert action (describing only the propagating spin-2 part of metric) versus more general actions that are non-linear in $R$. The latter are known as the $f(R)$-gravity actions, whose Lagrangian is given by 
\be \lb{act}
S = -\ha \int d^4 x \, \sqrt{-g}\, f(R)~,
\ee
and they represent the particular class of modified gravity theories which can provide 
viable {\it geometrical} description of inflation as well as dark energy (see eg. Ref.~\cite{tsu} for a review), in agreement with all known physical observations. The
current status of the $f(R)$ gravity theories is phenomenological (or macroscopic) and
truly non-perturbative (or non-linear).

In the context of the $f(R)$ gravity models of inflation the action (1) is the simplest representative of the class of viable actions (\ref{act}), whose function $f$ takes the form
$f=R- \fracmm{R^2}{6M^2}A(R)$ in the high-curvature regime with the slowly varying function $A(R)$ subject to the conditions
\be A(0)=1~,\qquad  |A'(R)|\ll \fracmm{A(R)}{R}~,\qquad  |A′′''R)| \ll \fracmm{A(R)}{R^2}~,
\ee
where the primes denote the derivatives with respect to $R$. The $R^2$ term dominates during inflation,
while the coefficient in front of the $R^2$-action is dimensionless. It gives rise to the
scaling invariance of the Starobinsky inflation in the large $R$ limit. This scaling invariance is not exact for finite (large) values of $R$, and its violation is exactly measured by the slow-roll parameters, in full correspondence to the nearly conformal spectrum of the CMB perturbations presumably associated with the scalaron field. It was proposed in Ref.~\cite{kstar2} to identify the scalaron with the Goldstone boson associated with the spontaneously broken scale invariance.

The viable $f(R)$ gravity models of the (dynamical) dark energy have very complicated 
$f(R)$-functions (see eg. Refs.~\cite{ab,hs,stard}) in order to obey the Newtonian limit, 
the Solar System tests and the cosmological tests via the Chameleon mechanism \cite{kh}.

Hence, in order to properly supersymmetrize $f(R)$ gravity to the form suitable for its
viable applications to inflation, reheating and dark energy, we need a generic $N=1$ 
locally supersymmetric action subject to the following neccessary conditions:
\begin{itemize}
\item it should depend only upon the single supergravity superfield having the scalar curvature
$R$ amongst its field components,
\item it should be possible to dualize that action to the standard matter-coupled
supergravity without higher derivatives.
\end{itemize}
The first condition guarantees the (super)gravitational origin of the theory.  The second condition is needed to avoid ghosts. In this Letter we propose such action in the old-minimal Poincar\'e supergravity for describing the Starobinsky inflation and the subsequent reheating.

In Sec.~2 we briefly review our setup. In Sec.~3 we propose the action subject to the conditions formulated above. Sec.~4 is our Conclusion. 

\section{Old-minimal Supergravity Setup}\label{sec:Setup}

We use the old-minimal formulation of supergravity in curved superspace (see eg. Refs.~\cite{a1,a2,a3,theisen} for details), which is manifestly supersymmetric. We use the lower case middle Greek  letters $\m,\n,\ldots=0,1,2,3$ for curved spacetime vector indices, the  lower case early Latin letters $a,b,\ldots=0,1,2,3$ for flat (target) space 
vector indices, and the lower case early Greek letters $\a,\b,\ldots=1,2$ for chiral spinor indices. The flat superspace indices together are denoted by capital early Latin letters, the curved superspace indices together are
denoted by capital middle Latin letters.

The supergravity superfield $\car$ containing the Ricci scalar curvature $R$ amongst its field 
components is covariantly chiral, $\Bar{\nabla}_{\dt{\a}}\car=0$, and obeys other off-shell constraints
 \cite{a1,a2,a3}. In our notation \cite{myrev} its bosonic field components are
\be \lb{sc}
\left. \car\right| = X,~\qquad \left. \nabla_{\a}\nabla_{\b} \car \right| = \ha \ve_{\a\b}
\left( -\frac{1}{3}R + 16 \bar{X}X +\frac{2}{9}b_ab^a +\ldots\right),
\ee
where the $(\nabla\low{\a},\Bar{\nabla}_{\dt{\a}})$ are the spinor covariant derivatives in curved superspace of the old-minimal supergravity, the vertical bars denotes the leading field components of the superfields, the horizontal bars denote Hermitian conjugation, and the dots stand for the fermionic contributions.  The complex scalar $X$ and the real vector $b_a$ are the standard set of the "auxiliary" fields in the old-minimal off-shell formulation of supergravity. However, in the context of field theory with higher derivatives, they can become propagating. Since it may lead to ghosts, demanding the ghost
freedom is one of the reasons for the second condition at the end of our first Section.

In the old-minimal formulation of supergravity (after the superconformal gauge fixing with a chiral
compensator) there exist an invariant chiral superspace, where the chiral compensator is converted into
the chiral density $\ce$ whose bosonic field components are
\be \lb{cd}
\left. \ce \right| =\ha e~,\qquad \left. \nabla^2 \ce \right| = -12e\Bar{X} +\ldots,
\ee
 where $e=\sqrt{-g}$ is the spacetime density. Therefore, one can define the invariant chiral superspace action \cite{gket}
\be \lb{fact}
 S_F = \int d^4xd^2\Theta\,2\ce F(\car) + {\rm H.c.}~~,
\ee
in terms of an arbitrary holomorphic potential $F$. The standard (textbook) Poincar\'e supergravity corresponds to
the choice $F(\car)=-6\car$. The action (\ref{fact}) was proposed in Ref.~\cite{gket} (where it was dubbed the $F(\car)$ supergravity) as a supersymmetric extension of the $f(R)$ gravity action (\ref{act}). Its cosmological applications were studied in Refs.~\cite{ket2,gkn,ket3,ket4,kwata,2kw,kstar,ktsu,yoko}. There exist a more general invariant {\it chiral} action with the potenial $F(\car,W^2)$ depending
upon the covariantly chiral Weyl superfield $W\low{\a\b\g}$, $\Bar{\nabla}_{\dt{\a}}W\low{\a\b\g}=0$, that has the
Weyl tensor amongst its field components \cite{ter}.

Besides having the manifest local $N=1$ supersymmetry, the action (\ref{fact}) also has the so-called {\it auxiliary freedom} \cite{g12} because the scalar auxiliary field $X$ does not propagate. The scalar curvature $R$ 
enters  the action (\ref{fact}) linearly but has a non-minimal coupling to $X$. Therefore, the algebraic equation of motion for $X$ can be solved for $X$ as a function of $R$. Substituting the solution back into the action yields some function $f(R)$ in the Lagrangian with the fermionic extension required by supersymmetry. Unfortunately, the class of real functions $f(R)$ that can be obtained by this procedure from a holomorphic (or analytic) master function $F$ 
appears to be too narrow for describing a viable inflation.

However, the requirement of the auxiliary freedom is not necessary as long as it does not lead to ghosts.  The 
ghost freedom is also achieved provided that the higher derivatives can be eliminated by a duality transformation. Then more general supersymmetric actions are possible, though with some extra physical degrees of freedom. 
A general pattern for those actions was found the long time ago by Cecotti \cite{cec} by using the $N=1$ superconformal tensor calculus, and it reads~\footnote{In our approach the superconformal tensor 
calculus is not necessary.}
\be \lb{non2}
 S_N = \int d^4 xd^4\theta\, E^{-1} N(\car, \Bar{\car})
\ee
in terms of a non-holomorphic potential $N(\car, \Bar{\car})$. The integration in Eq.~(\ref{non2}) goes 
over the whole curved superspace with a supervielbein $E_A{}^M(x,\theta,\Bar{\theta})$ and its superdeterminant (Berezinian) $E={\rm sdet}(E_A{}^M)={\rm Ber}(E_A{}^M)$. It is obvious that the supergravity "auxiliary" field $X=\left.\car\right|$ is propagating in the theory (\ref{non2}), while the form of Eq.~(\ref{non2}) is similar to a
generic kinetic term of a chiral superfield. Of course, in the standard (textbook) supergravity, we have 
$N(\car, \Bar{\car})=-3=const$.

\section{Action}\label{sec:Action}

The new action we propose is just a sum of Eqs.~(\ref{fact}) and (\ref{non2}),
\be \lb{action}
S= \int d^4 xd^4\theta\, E^{-1} N(\car, \Bar{\car}) +\left[ \int d^4xd^2\Theta\,2\ce F(\car) + {\rm H.c.} \right]~, 
\ee 
and it is parametrized by two arbitrary functions, a non-holomorphic (real) potential $N(\car, \Bar{\car})$
and a holomorphic (analytic) potential $F(\car)$. The action (\ref{action}) is similar to a generic action of an independent dynamical chiral matter superfield $\F$ minimally coupled to supergravity (without higher derivatives), having the form
\be \lb{genm}
S= -3\int d^4 xd^4\theta\, E^{-1} e^{-\frac{1}{3}K}+\left[ \int d^4xd^2\Theta\,2\ce W + {\rm H.c.} \right]~, 
\ee
with the K\"ahler potential $K(\F,\Bar{\F})$ and the superpotential $W(\F)$. However, the chiral superfield $\car$ is not independent but is given by the constrained chiral scalar curvature superfield, so that the action (\ref{action}) has higher derivatives. The scalar curvature $R$ enters the action (\ref{action}) quadratically, but with
non-minimal couplings of $X$ to both $R$ and $R^2$.

In order to understand the physical significance of the action (\ref{action}), let us rewrite it to the form
\be \lb{action2}
\eqalign{
 S~= &~ \int d^4 xd^4\theta\, E^{-1} N(J, \Bar{J}) +\left[ \int d^4xd^2\Theta\,2\ce \L(J-\car) + {\rm H.c.} \right] \cr
 &~ + \left[ \int d^4xd^2\Theta\,2\ce\left( -\cy\car +Z(\cy)\right) + {\rm H.c.} \right]~, \cr}
\ee 
where we have introduced two independent (covariantly) chiral superfields, $J$ and $\cy$, the 
covariantly chiral Lagrange multiplier superfield $\L$, and the holomorphic function $Z(\cy)$ which is
 the Legendre transform of $F(\car)$, 
\be \lb{let1}
 Z(\cy) = F(\car(\cy)) +\cy\car(\cy)~,
\ee
where the arguments $\car$ and $\cy$ are algebraically related by the equations
\be \lb{let2}
 \cy = -F'(\car)\qquad {\rm and} \qquad \car =Z'(\cy)~.
\ee

The algebraic equation of motion of the superfield $\cy$, which follows from the 
variation of the action (\ref{action2}) with respect to $\cy$, together with its solution
coincides with Eq.~(\ref{let2}). Substituting the solution into the action (\ref{action2})
yields back the holomorphic terms in Eq.~(\ref{action}) because of Eq.~(\ref{let1}).
Varying the action (\ref{action2}) with respect to $\L$ yields
\be \lb{leq} J=\car  
\ee
so that its substitution into the action (\ref{action2}) gives back the non-holomprphic term in
Eq.~(\ref{action}).  It proves the classical equivalence of the actions (\ref{action}) and (\ref{action2}). 

The action (\ref{action2}) does not have higher derivatives and can be transformed into the standard form of the matter-coupled supergravity theory, as is shown below.

First, by using the (Siegel's) identity in supergravity, the action (\ref{action2}) can be rewritten to
\be \lb{action3} 
\eqalign{
 S~= &~ \int d^4 xd^4\theta\, E^{-1}\left[ N(J, \Bar{J}) -(\cy+\Bar{\cy}) -(\L+\Bar{\L})\right] \cr
&~ + \left[ \int d^4xd^2\Theta\,2\ce\left( Z(\cy) +\L J\right) + {\rm H.c.} \right] \cr}
\ee 
After a change of variables, $\ct=\cy+\L$, it is easy to see that the superfield $\cy$ does not enter
the kinetic terms in the action (\ref{action3}) so that its equation of motion is algebraic, and it reads
\be \lb{emcy}
Z'(\cy) =J~.
\ee
The solution to Eq.~(\ref{emcy}) is therefore given by
\be \lb{scy}
\cy = -F'(J)
\ee
because of Eqs.~(\ref{let1}) and (\ref{let2}).

We are now in a position to read off the K\"ahler potential $K(J,\Bar{J};\ct,\Bar{\ct})$ 
and the superpotential $W(\ct,J)$ of the matter-coupled supergravity theory in Eq.~(\ref{action3})
in terms of two dynamical chiral superfields $\ct$ and $J$ as follows:
\be \lb{kal}
K = -3 \ln \left[ \fracmm{ \ct+\Bar{\ct} - N(J,\Bar{J}) }{3}\right]
\ee
and
\be \lb{sup}
W = Z(-F'(J)) + J\left[ \ct + F'(J)\right] ~.
\ee

The bosonic sector of this theory is given by \cite{crem}
\be \lb{bosa}
e^{-1} L_{\rm bos.} = -\ha R + K_{i,\bar{j}} \pa_{\m} z^i\pa^{\m} \bar{z}^{\bar{j}} -V
\ee
with the scalar potential \cite{crem}
\be \lb{spot}
V(z,\bar{z}) = e^G \left[ G_{,i} \left( \fracmm{\pa^2 G}{\pa z^i\pa\bar{z}^{\bar{j}}} \right)^{-1}
G_{,\bar{j}} -3 \right]
\ee
where we have introduced the notation
\be \lb{not2}
\left. \ct\right|=z^1 \qquad {\rm and} \qquad \left. J\right| =z^2~,\qquad i,j=1,2
\ee
and the function \cite{crem}
\be \lb{mf}
G = K +\ln(\Bar{W}W)~,\qquad G_{,i}=\fracmm{\pa G}{\pa z^i}~,\qquad G_{,\bar{j}}=
\fracmm{\pa G}{\pa\bar{z}^{\bar{j}}}~.
\ee

The K\"ahler potential in Eq.~(\ref{kal}) is of the {\it no-scale} type \cite{nos}. In particular, when
$N(\bar{J},J)=\Bar{J}J$, it describes the non-linear sigma-model \cite{myb} on the non-compact
homogeneous space $SU(2,1)/SU(2)\times U(1)$.

The special cases of the theory (\ref{action}) in its dual form above were extensively studied in the recent literature devoted to embedding of the Starobinsky inflation into supergravity. The case of $N=0$ with an
arbitrary $F$ was investigated in Refs.~\cite{myrev,ktsu}. The case of  $N=N(\Bar{J}J)$ with $F'(J)=-1/2$ was used in Refs.~\cite{kl1,kl2,kl3,kl4} to generate the inflationary scalar potential (\ref{starp})  in the parametrization $\ct=\ha \exp\[\sqrt{\frac{2}{3}}\f] +ib$ along the inflationary trajectory $J=b=0$. The need to stabilize that inflationary trajectory requires a particular choice of the function $N(\Bar{J}J) =\Bar{J}J - \z (\Bar{J}J)^2$ for appropriate values of the positive parameter $\z$ \cite{kl4}. 

Our theory in the dual form defined by Eqs.~(\ref{kal}) and (\ref{sup}) also overlaps with the $SU(2,1)/U(2)$ no-scale supergravity models studied in Ref.~\cite{el1} with a generic superpotential $W(\ct,J)$ and $N=\Bar{J}J$. The specific superpotentials $W$ reproducing the scalar potential (\ref{starp}) were proposed in Ref.~\cite{el1} too.~\footnote{See Ref.~\cite{hi} for the earlier models of that type.} Stability of the inflationary trajectory in the approach \cite{el1} can also be achieved by adding a quartic term $(\Bar{J}J)^2$ (with a negative coefficient $<17$) to the naive $N$-function equal to $\Bar{J}J$ in the K\"ahler potential.    

In our approach both superfields $J$ and $\ct$ have the supergravitational origin, being related to the single chiral supergravity superfield $\car$ in the original picture (with higher derivatives), while the superpotential (\ref{sup}) has 
the particular (not arbitrary) form.

\section{Conclusion}

Our main results are given by Eqs.~(\ref{action}), (\ref{action3}), (\ref{scy}), (\ref{kal}) and (\ref{sup}). Most of those
actions can be extracted from the Cecotti's paper \cite{cec} after the superconformal gauge fixing. In this Letter we
identified those old-minimal Poincar\'e supergravities that are needed for realization of the Starobinsky inflation (and, perhaps, the present dark energy too) in supergravity, when using only a single supergravity superfield $\car$.

The alternative to the old-minimal formulation of supergravity is provided by its new-minimal formulation
\cite{a1,a2,a3}  where (in the superconformal context relating the two versions) the chiral compensator is
replaced by the real linear compensator. Those off-shell formulations are inequivalent in the supergravity theories with higher derivatives. As was demonstrated in Ref.~\cite{gre}, the new-minimal version of supergravity can be successfully used for embedding the scalar potential (\ref{starp}), while the dual version
of that theory has a dynamical massive vector multiplet whose loweset real scalar field component plays the role of inflaton. The approach of Ref.~\cite{gre} was extended in Ref.~\cite{fklp}, where it was found that the general master action of the new-minimal supergravity is governed by a single real potential, so that any single-real-field scalar potential, which is the square of a real function (related to the real potential), can be  upgraded to the old-minimal supergravity. That approach is not directly related to $f(R)$ gravity.
Our master action (\ref{action}) has two potentials, one holomorphic and another non-holomorphic, so that
it seems to be more powerful.

Quantum corrections to the low-energy effective action (\ref{action}) may appear in the form of the higher order
terms with the covariant derivatives of the scalar curvature superfields $\car$ and $\Bar{\car}$. As was shown
in Ref~\cite{gre}, those corrections may destabilize the Starobinsky inflation by reducing the size of the plateau in the scalar potential (\ref{starp})  -- the plateau should have the size about 5 in the Planck units, in order
to achieve $N_e>50$. It is not very surprising from the point of view of $f(R)$ gravity -- it is well known that the higher order terms in the scalar curvature (beyond the $R^2$) represent a threat to the Starobinsky inflationary  scenario (see eg. Ref.~\cite{myrev}).

Of course, our main purpose is not just to demonstrate that the Starobinsky inflation is compatible with local supersymmetry, and it is much more than embedding of a particular scalar potential into supergravity. The
natural appearance of the no-scale supergravity from the action (\ref{action}) signals good perspectives for
unifying inflation with particle physics (beyond the Standard Model) because the no-scale supergravity can be
used for a dynamical solution to the hierarchy between the electro-weak and Planck scales. And the no-scale
supergravity naturally emerges as the low-energy effective action in superstring compactifications.

\section*{Acknowledgements}

The author thanks Alex Kehagias for discussions. This work was supported by the World Premier International Research Center Initiative (WPI Initiative), MEXT, Japan, the Japanese Society for Promotion
of Science (JSPS), and the Research Council of Norway (RCN).

\end{document}